# Optimized substrates and measurement approaches for Raman spectroscopy of graphene nanoribbons

*Jan Overbeck*[*,1,2,3,#], *Gabriela Borin Barin*[*,1,#], *Colin Daniels*[4], *Mickael Perrin*[1], *Liangbo Liang*[5], *Oliver Braun*[1,2], *Rimah Darawish*[1,6], *Bryanna Burkhardt*[4], *Tim Dumslaff*[7], *Xiao-Ye Wang*[7,†], *Akimitsu Narita*[7], *Klaus Müllen*[7,8], *Vincent Meunier*[4], *Roman Fasel*[1,6], *Michel Calame*[1,2,3] *and Pascal Ruffieux*[*,1]

[1] Empa, Swiss Federal Laboratories for Materials Science and Technology, 8600 Dübendorf, Switzerland
[2] University of Basel, Department of Physics, 4056 Basel, Switzerland
[3] University of Basel, Swiss Nanoscience Institute, 4056 Basel, Switzerland
[4] Rensselaer Polytechnic Institute, Department of Physics, Applied Physics, and Astronomy, Troy, New York 12180, United States
[5] Oak Ridge National Laboratory, Center for Nanophase Materials Sciences, Oak Ridge, Tennessee 37831, United States
[6] University of Bern, Department of Chemistry and Biochemistry, 3012 Bern, Switzerland
[7] Max Planck Institute for Polymer Research, Ackermannweg 10, 55128 Mainz, Germany
[8] Johannes Gutenberg-Universität Mainz, Institute of Physical Chemistry, 55128 Mainz, Germany

[†] current address: State Key Laboratory of Elemento-Organic Chemistry, College of Chemistry, Nankai University, Tianjin 300071, China

[#] These authors contributed equally

*corresponding authors: gabriela.borin-barin@empa.ch
jan.overbeck@empa.ch
pascal.ruffieux@empa.ch

**Abstract**

The on-surface synthesis of graphene nanoribbons (GNRs) allows for the fabrication of atomically precise narrow GNRs. Despite their exceptional properties which can be tuned by ribbon width and edge structure, significant challenges remain for GNR processing and characterization. In this contribution, we use Raman spectroscopy to characterize different types of GNRs on their growth substrate and to track their quality upon substrate transfer. We present a Raman-optimized (RO) device substrate and an optimized mapping approach that allows for the acquisition of high-resolution Raman spectra, achieving enhancement factors as high as 120 with respect to signals measured on standard $SiO_2$/Si substrates. We show that this approach is well-suited to routinely monitor the geometry-dependent low-frequency modes of GNRs. In particular, we track the radial breathing-like mode (RBLM) and the shear-like mode (SLM) for 5-, 7- and 9-atom wide armchair GNRs (AGNRs) and compare their frequencies with first-principles calculations.

**Introduction**

Graphene nanoribbons (GNRs) - narrow stripes of graphene – have unique electronic properties that make them an interesting material for nanoelectronic devices. In contrast to graphene, GNRs have a sizeable bandgap due to quantum confinement, which is a fundamental requirement for room temperature switching applications[1,2]. The electronic properties of GNRs can be tuned by engineering their width and edge structure[3,4]. Specifically, armchair GNRs (AGNRs) show a width-dependent electronic bandgap[5]. According to their width expressed in units of carbon atoms (N) across the ribbon, N-AGNRs can be classified into the three families N = 3p (medium bandgap), 3p+1 (wide bandgap) and 3p+2 (quasi-metallic), where p is an integer. Within each family, the bandgap scales inversely with GNR width[6]. Recent advances in the on-surface synthesis of GNRs have allowed to reach the required selectivity and atomic control over width and edge structure[7,8]. Scanning probe microscopy and spectroscopy studies have confirmed the intimate structure-property relationship by providing morphological and electronic information at the atomic-scale[1,4,9,10]. The on-surface synthesis of atomically precise GNRs is, however, just a first step towards integrating GNRs into nanoelectronic devices which needs to be followed by their controlled transfer from the metallic growth substrate (usually Au(111)) onto an insulating or semiconducting substrate appropriate for digital logic applications [11,12]. In view of device integration, this is a critical step since the quality of the GNRs needs to be preserved and thus monitored after substrate transfer. Raman spectroscopy is, so far, the only technique able to probe the structural quality of GNRs all the way from growth under ultra-high vacuum (UHV) conditions to device integration [5,13,14]. This was demonstrated in a study reporting the first field-effect transistors (FETs) with large on/off ratio relying on 9-AGNRs as the channel material[13]. In particular, Raman spectra before and after GNR transfer were compared and the devices' high performance was directly linked to the presence of the RBLM on the device. So far, however, this approach has been limited to cases with good resonance enhancement. Moreover, after transfer onto a silicon-based device substrate most of the GNRs' low-frequency modes are hidden in the silicon background. As these low-frequency modes typically have low intensities, they require high laser powers and/or long integration times to be detected, which adversely affect the structural integrity of the GNRs.

Here, we report on the fabrication of Raman-optimized (RO) device substrates relying on the interference-based intensity enhancement provided by an amorphous dielectric layer on a metal which blocks the background of the silicon underneath. Together with an advanced mapping approach, this results in high signal-to-noise (S/N) ratios for several excitation wavelengths while limiting radiation damage to the GNRs under investigation. Importantly, the RO layer is integrated into the device substrate itself, allowing us to systematically probe the GNRs' quality by investigating their low-frequency Raman modes. We apply this procedure to three different ribbons (5-, 7-, and 9-AGNRs) that cover all the AGNR families and compare Raman spectra obtained directly on the gold growth substrate with those on the RO-substrate. Finally, we discuss in detail the low-frequency modes resolved in this way and compare them with first-principles calculations.

**Results and discussions**

Raman spectroscopy has been widely used to characterize graphite and carbon-based nano-materials over the last five decades[15]. It is fast, potentially damage-free, and particularly well suited to investigate the morphology of carbon materials at the nanoscale[16,17]. GNRs have fingerprint modes that are used to probe their quality upon growth and integration into devices[12]. The Raman spectrum of GNRs is dominated by a strong mode around 1600 $cm^{-1}$, known as the G mode, which is common to all $sp^2$ carbon materials and is assigned to the in-plane optical vibrations of the carbon $sp^2$ lattice. The high-frequency region of the spectrum also shows modes between 1100-1400 $cm^{-1}$ which are a signatures of the presence of hydrogen at GNRs' passivated edges[18]. The low-frequency part of the spectrum is dominated by the radial breathing-like mode (RBLM) with a frequency that scales with ribbon width[19,20]. The lowest reported frequency in the spectrum of GNRs is the shear-like mode (SLM) for which the atoms on the two sides of the ribbon move in opposite directions along the GNR-axis[21].

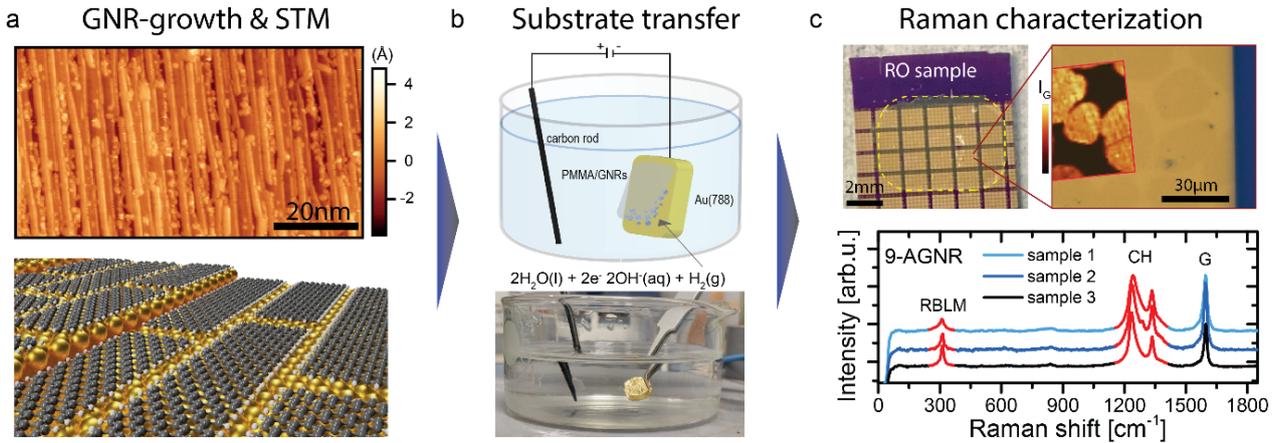

**Figure 1. Sample fabrication and characterization. a:** STM image of aligned 9-AGNRs (top) and a sketch of the ribbons grown parallel to the Au(788) terraces (bottom). **b:** Sketch of the electrochemical delamination transfer (top) and a picture of the on-going transfer (bottom). **c:** Image of a Raman-optimize (RO) device substrate with a transferred PMMA/GNR film on top (dashed outline). Optical zoom-in with Raman G-intensity map as overlay (top). Raman spectra of three different transfers illustrating the sample-to-sample variation of the RBLM and CH-spectral region (bottom).

Figure 1 shows an overview of the fabrication and characterization process of a GNR device from UHV-synthesis to the transfer onto a RO-device substrate (see experimental section for details). Graphene nanoribbons are grown under UHV conditions on a vicinal Au(788) surface[4], which promotes unidirectional growth on the narrow (111) terraces along the step edges, as depicted by the scanning tunneling microscope image in Figure 1a. Highly-aligned arrays of GNRs are important to control device properties and can significantly increase device yield as well as the drain current in the transistor "on" -state which is essential to meet the demands of switching applications[22]. In a second step, GNRs are transferred from the Au(788) growth substrate onto the RO-substrate using an electrochemical delamination technique (Figure 1b) [23]. This method preserves the uniaxial GNR alignment upon substrate transfer and thus allows for a well-defined orientation of the ribbons with respect to the device electrodes. We use Raman spectroscopy to investigate GNR quality and orientation after substrate transfer. In Figure 1c, we show an example for the characterization of a transferred GNR-film. Here, the good optical contrast on the RO-substrate and large area Raman mapping (overlay inset) reveal the bubble pattern of an inhomogeneously transferred film. Below, we show Raman spectra of 9-AGNRs from different transfers. These measurements performed on RO-substrates show significant sample-to-sample variations, as highlighted in the Raman profiles, underlining the importance of monitoring GNR quality after every process step. In the following, we will first discuss the mapping strategy and its benefits for the characterization of GNR-samples before describing the RO-substrates in more detail.

The usual approach to acquire high-quality Raman spectra is to extend the integration time. This approach, however, is problematic for GNRs because it leads to the introduction of defects by prolonged radiation exposure as has been previously reported by Senkovskiy *et al*. for the case of 7-AGNRs[23].

In contrast, scanning a large sample area allows us to ensure that we capture the typical properties of the GNRs, exclude outliers and get the best signal-to-noise spectra with minimal damage to the ribbons. We investigate this approach for the 5- and 9-AGNRs transferred to an RO-substrate but note that this is a general observation for GNR samples. Figure 2a shows a time series of Raman spectra acquired on a single point of a GNR film, showing a rapidly decaying intensity (over a measurement window of 100 seconds). Performing such measurements as a function of laser power (Figure 2b, top panel) shows that the damaging rate depends on light intensity. Importantly, it reveals that the damage cannot be fully avoided by reducing the laser power below a threshold. In fact, the signal intensity can be scaled to a constant power-integration time product, as shown in Figure 2b (bottom panel), indicating that the damage mechanism scales with the number of photons.

Performing the same measurements in vacuum or with an infrared laser strongly reduces this damaging behavior as highlighted in Figure 2b (blue and red profiles, respectively).

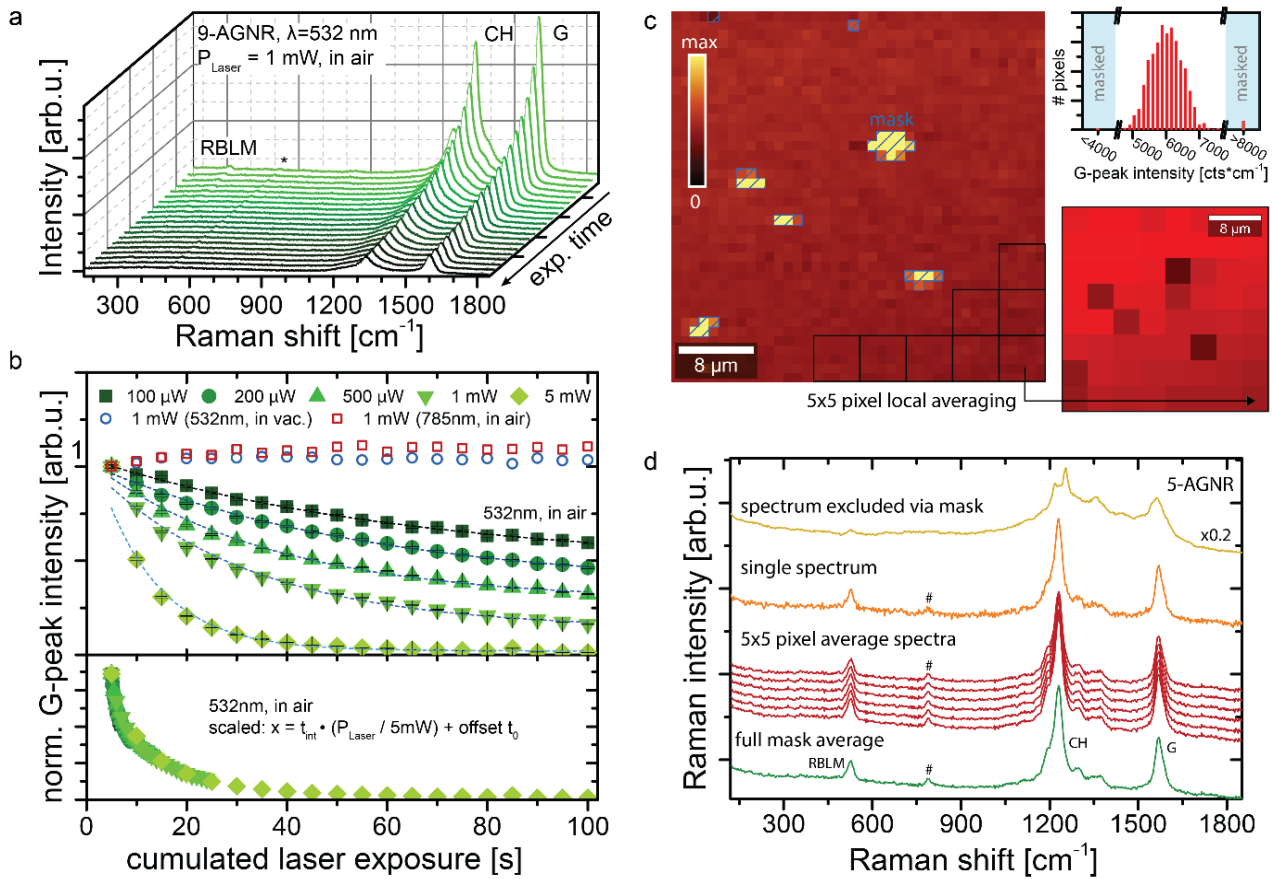

**Figure 2. Laser-induced damage of GNRs and Raman mapping. a:** Waterfall plot of a time series (cumulated laser exposure = 100 seconds) of Raman spectra on 9-AGNRs. The mode indicated by an asterisk is background from the Si-substrate. **b:** Top panel: G-peak intensities extracted from Lorentzian fits for several time series as a function of cumulated exposure time. The time dependence is fitted with an exponential decay for each power. For comparison, results from time series in vacuum and with NIR-excitation are shown in blue and red, respectively. Bottom panel: Signal intensity scaled to a constant power-integration time product. **c:** Raman map with 40x40 pixels of 5-AGNRs on Au/Mica. The histogram of G-peak intensities is shown in the top right and used to create a mask to exclude outliers. Local averaging is used to increase the signal-to-noise ratio and probe the homogeneity of less intense modes across the sample. **d:** Raman spectra from map displayed in c, top to bottom: impurity-dominated spectrum excluded *via* mask, single spectrum with S/N too low to clearly identify the mode labelled '#', local averages of 5x5 pixels each showing mode '#', full average over 1566 spectra in mask for best S/N ratio.

Even though measuring in vacuum avoids this damaging issue, it may not be available on every experimental Raman setup and substantially increases the time necessary for investigating a sample, thereby limiting the usefulness of Raman spectroscopy to monitor device fabrication. The second approach of exclusively using infrared excitation prevents wavelength-dependent studies and is limited to Raman modes that are (near-) resonantly enhanced at these wavelengths, as non-resonant Raman is often too weak to reveal several spectroscopic features of interest in these (sub-) monolayer systems. It is therefore important to limit the power-integration time product to values compatible with minimal GNR damaging.

In order to still achieve a good signal-to-noise ratio, we employ a large area mapping process (Figure 2c) in which a Raman map with hundreds of spectra with limited integration time is recorded. The sample is scanned at a constant speed during the acquisition of a map, such that it is displaced by typically one laser spot diameter per acquisition interval. If the integration time needed to get sufficient signal-to-noise is comparable to the previously determined timescale of ribbon damaging, the sample is scanned at a higher speed such that the radiation damage during one integration time gets spatially distributed. The resulting loss in spatial resolution is usually acceptable for homogeneous GNR-films. Typically, a map of the G-peak intensity is used to check for sample homogeneity. A histogram of the peak intensity is then used to identify outliers, which are excluded if their spectra display atypical signatures that are not a simple, moderate scaling of the spectrum (Figure 2c and 2d). If necessary, averaging of neighboring pixels is used to obtain a larger signal-to-noise ratio that allows to clearly identify peaks in the spectra representing a part of the scanned area. An example of this is the low intensity peak labelled # in Figure 2d, which could be mistaken for noise in a single spectrum. Local averaging reveals its homogeneous presence across the sample. Finally, an average spectrum is calculated from the entire homogenous sample area (excluding the mask) to give the best signal-to-noise ratio (Figure 2d).

In order to use Raman spectroscopy as a tool to assess GNR quality and stability, which are both critical for applications, it is desirable to perform Raman measurements directly on the final device. This is often hampered by the presence of a significant background due to the substrate (usually $SiO_2$/Si), which masks the GNR-related modes of interest. We illustrate this behavior in Figure 3a where we compare the Raman spectra of 5-, 7- and 9-AGNRs with that of a $SiO_2$/Si-substrate. The Raman spectrum of silicon consists of two strong optical phonon peaks at 520 $cm^{-1}$ and ~950 $cm^{-1}$, and an acoustic phonon peak at 300 $cm^{-1}$ [24]. These three modes are in the same spectral region as the acoustic GNR-modes (RBLM and SLM) that are most useful for characterizing geometry-dependent properties of GNRs. For the three different ribbons investigated in this study, we observe the RBLM at 529, 398 and 311 $cm^{-1}$ for 5-, 7- and 9-AGNRs, respectively. In Figure 3a we highlight in red the spectral regions for which the Si-background at an excitation wavelength $\lambda_{ex}$ = 488 nm masks the signal in a representative sample of 9-AGNRs.

To address this issue we developed a layered, interference-optimized substrate that is suitable for both Raman and transport measurements. Interference enhanced Raman scattering takes into account that the layer structure of the substrate plays an important role in the measured Raman intensity, in addition to the usual factors such as set laser power, scattering cross-section of the investigated material, and numerical aperture of the objective[25]. This is particularly clear for the interference of incoming and reflected laser beam that results in very different effective electromagnetic fields and therefore Raman intensities for GNRs on metallic or insulating substrates[26]. Blake and coworkers also reported on the optimization of substrate layer-thicknesses to improve optical contrast for the fabrication of graphene-based devices by taking light interference into account[27]. Interference models, have further been used to describe the overall Raman intensity of thin films including graphene on oxides and explain the change in relative peak intensities as a function of oxide thickness[26,28–30]. Here, we combined these considerations to produce a substrate that is compatible with standard silicon-based fabrication approaches for nanoelectronic devices, yields good contrast for GNR-film visibility and leads to an enhanced Raman signal of GNRs without the otherwise dominant signal of silicon as a background.

When measuring GNRs directly on an Au(788) surface or on a metallic contact pad of a device used for fabrication of GNR-based transistors, destructive interference leads to a low electric field (and therefore low Raman scattering intensity) experienced by the GNR layer (see Supporting Information for a detailed discussion and simulations). Qualitatively, one can think of the metal as a quasi-perfect electric conductor that imposes an electric field node as a boundary condition. For a real metal with finite skin-depth this can be partially overcome by using higher laser powers to obtain a signal at all (which in this case is possible without excessive GNR damaging). A common approach to overcome this issue is surface-enhanced Raman scattering on a nanostructured metallic substrate. In our case, however, this is not desired, as we want to assess the average properties of the GNR film and not those in plasmonic hot-spots. Transferring the ribbons to a silicon substrate, instead, results in the background problems described above. Alternatives also include using an amorphous Raman substrate such as $CaF_2$, resulting in a clean, low background signal, but at the cost of severely limiting the processability for devices.

The Raman-optimized (RO) structure we designed and fabricated is displayed in Figure 3b and consists of an atomic layer deposition (ALD)-grown oxide layer patterned on optically thick (typically 80-90 nm) metal source-drain contacting-pads (labelled S/D) on a silicon device-substrate, that acts as a support and optional gate (G). The result is enhanced optical visibility of the GNRs on top of the metal, allowing for easy identification of film-inhomogeneity and a strongly enhanced Raman intensity. This is a result of the GNR-layer being into the region of a field anti-node and allows the spectrum acquisition at much lower excitation power.

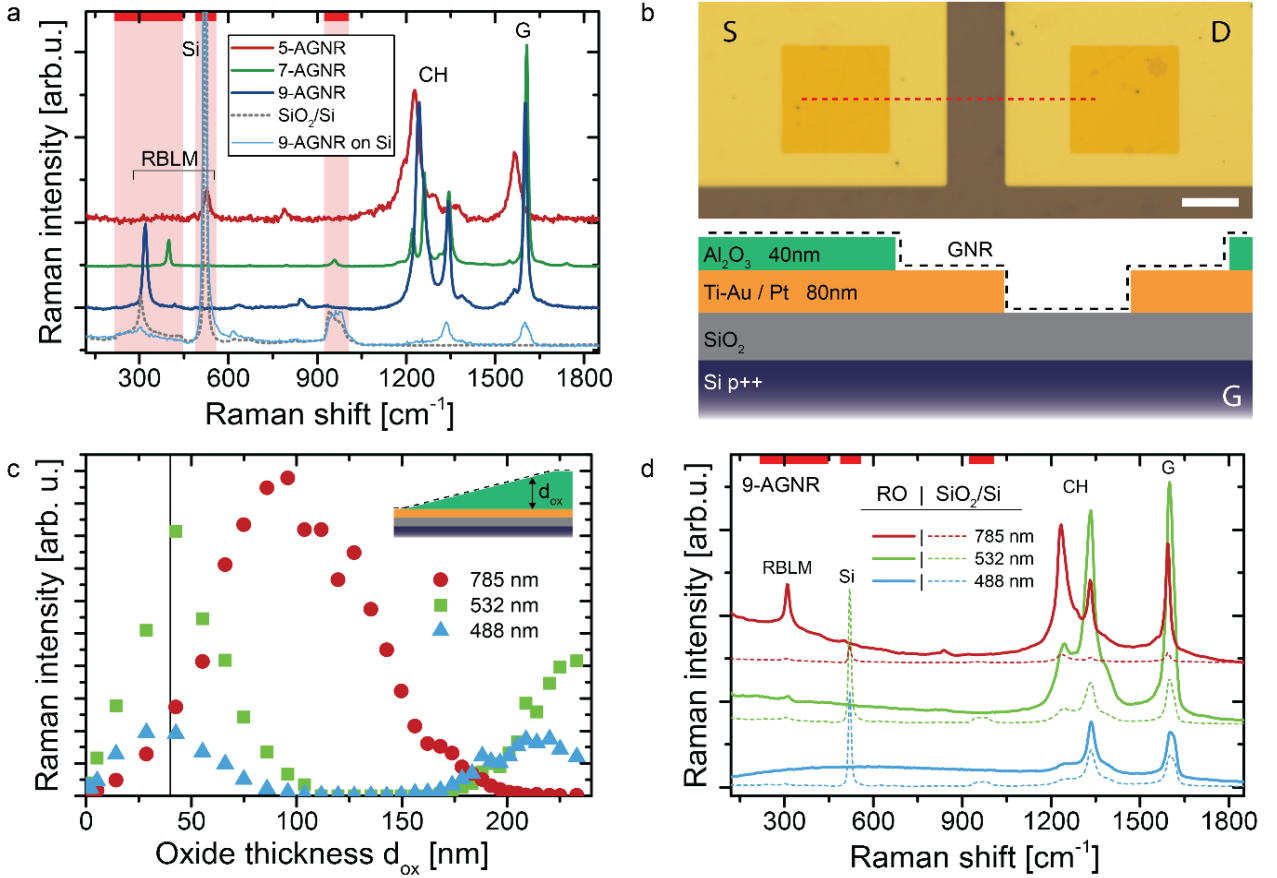

**Figure 3. Raman-optimized (RO) device substrates. a:** Raman spectra of 5-, 7- and 9-AGNRs on Au ($\lambda_{ex}$ = 785/532/785 nm, arbitrarily scaled) compared to SiO2/Si ($\lambda_{ex}$ = 785 nm) and 9-AGNR transferred to SiO2/Si ($\lambda_{ex}$ = 488 nm), scaled to the second order Si-peak. **b:** Optical micrograph of an RO-device substrate based on standard p-doped silicon with thermal oxide. The layers are sketched in the lower half of the panel, corresponding to the dashed red line in the optical image. **c:** Raman intensity of the 9-AGNR G-peak on top of an RO-substrate as a function of oxide thickness on a sample with an oxide gradient. **d:** Raman spectra of transferred 9-AGNRs measured in the interference-optimized region (RO, solid lines) and on adjacent SiO2/Si (dashed lines) with different wavelengths. Measured with 100x (NA=0.9), in air, no background subtraction.

In Figure 3c we show the Raman intensity of the G mode as a function of oxide thickness for a 9-AGNR sample measured with three different excitation wavelengths. For each wavelength, there is an optimal thickness resulting in maximum Raman intensity (see supporting Note 2). A good compromise suitable for multi-wavelength investigations of GNRs is found at an oxide thickness of about 40 nm (indicated by a vertical line), which was chosen for all subsequent studies. Figure 3d provides a comparison of the Raman spectra of 9-AGNRs transferred onto a 40 nm RO-substrate and the adjacent SiO2/Si substrate measured with $\lambda_{ex}$ = 785, 532 and 488 nm. The spectra are shown without any background subtraction and are normalized for power and integration-time. On the RO-substrate, the Si-background is suppressed and an enhancement of the GNR signal by a factor of 1.5/5.3/11.7 is observed for excitation wavelengths of 488/532/785 nm, respectively. Compared to the signal directly on the Au pad, we find an enhancement of 11.5/19.5/43.0, respectively.

Note, that these values are for an oxide thickness of 40 nm, which is a compromise between the wavelengths used in this study and constraints from sample fabrication. In Figure S2 we show that a substrate can be optimized for a particular wavelength of interest, by choosing an oxide thickness which satisfies the well-known condition of interference $d_{ox} \cdot n_{ox} = (2m+1) \cdot \lambda/4$, $m \in \mathbb{N}$, where $d_{ox}$ and $n_{ox}$ are the thickness and the refractive index of the oxide layer, respectively[30]. In this way, one can achieve enhancement factors as high as 120 with respect to a standard $SiO_2$/Si substrate. Moreover, significant enhancement is still possible by using high-quality thin gate-oxides as found in state-of-the-art field-effect transistors.

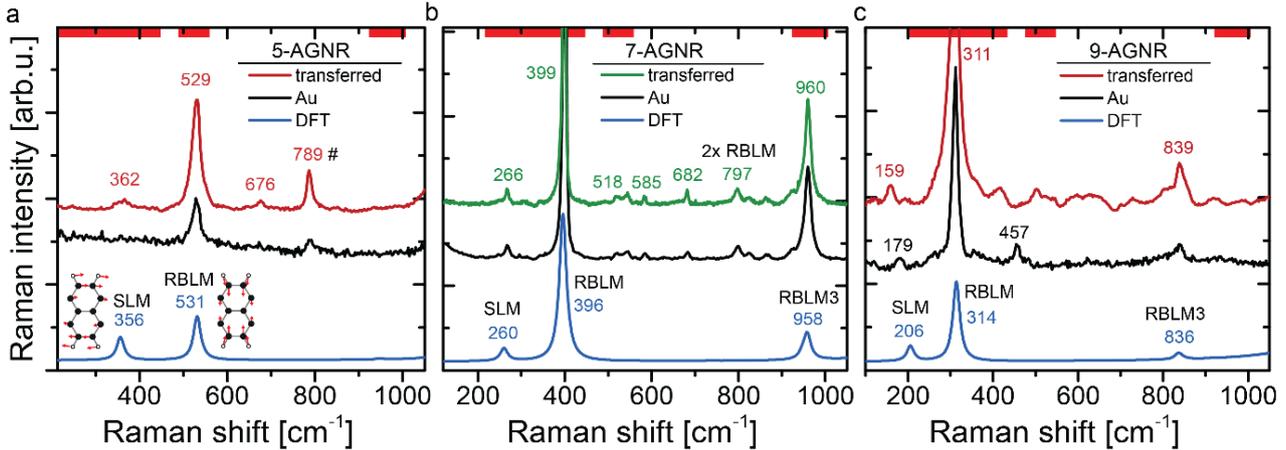

**Figure 4. Raman spectra before and after substrate transfer and comparison with DFT. a:** Low energy spectrum of 5-AGNRs ($\lambda_{ex}$ = 785 nm), **b:** 7-AGNRs (532 nm), and **c:** 9-AGNRs (785 nm) before and after transfer. Theoretical spectra were obtained by summing DFT-based resonant Raman intensities (all 785 nm excitation, Lorentzian line shapes with width of 10 cm$^{-1}$, no frequency scaling). The most prominent peaks are labelled with position and normal mode assignment. The normal mode displacements for the SLM and RBLM are exemplified for the 5-AGNR.

To explore the benefits of our RO-substrate and mapping strategy for investigating GNRs we acquire Raman spectra with high signal-to-noise ratio for 5-, 7- and 9-AGNRs before and after substrate transfer (Figure 4). The spectra acquired on the Au growth-substrate are displayed in black. The strong RBLM mode is visible for all three investigated ribbons. We also observe higher energy modes that are relatively clear in the case of 7-AGNRs, because this ribbon is resonant with 532 nm excitation. Similar features are much fainter for the less resonantly excited spectra of 5- and 9-AGNRs ($\lambda_{ex}$ = 785 nm). The spectra after transfer are shown in red and green. The peak position and width of the intense RBLM signal has been used to monitor GNR quality and to probe their stability over time[11,12]. Here, we observe several additional modes that we attribute to overtones such as the RBLM3 and higher order processes by comparison with computed normal modes (see Supporting Note 3). Most notably, for all GNRs we observe a mode at a frequency below the RBLM. This additional fingerprint of GNRs is the SLM, which is expected to have a similar scaling with GNR-width as the RBLM and has so far only been reported for 7-AGNRs[21]. Here, we can resolve this mode for all investigated ribbons, with frequencies of 362, 266 and 160 cm$^{-1}$ for 5- ,7- and 9-AGNRs, respectively.

Finally, we performed DFT-calculations and computed the Raman intensities using the VASP[31–33] and Phonopy software[34]. The calculated spectra are shown in blue and reproduce the experimental observations well. In particular, the calculated values for the SLM-frequency of 5- and 7-AGNRs closely match the experimental observations. For 9-AGNRs, there is some discrepancy with the calculations showing a mode frequency of 206 cm$^{-1}$, whereas experimentally we observe 179 cm$^{-1}$ on the Au growth substrate and 159 cm$^{-1}$ after transfer (see Supporting Note 3 for additional spectra). Periodic DFT, however, provides a highly idealized picture, ignoring a number of effects such as the presence of a substrate and defects in GNRs. In particular, 9-AGNRs are known to exhibit a substantial amount of phenyl defects which occur between the polymerization and the cyclodehydrogenation reaction[9]. Also, the structure of the precursor molecule for the 9-AGNRs results in slanted ribbon termini. We investigated the possibility of the edge-defects or the ribbon termini causing a mode softening that could account for the difference between experimental and DFT-calculated spectra. A preliminary analysis based on calculations with the Phonopy package suggests, however, that these features alone cannot account for a mode softening on the order of 20-40 cm$^{-1}$.

Finally, we note that our experimental spectra for 5-AGNRs reveal two peaks at 676 cm$^{-1}$ and 789 cm$^{-1}$ that are not present in the calculated spectra. The homogeneous presence of the second of these modes across the sample was already discussed in Figure 2, and it is found for several different GNR samples (Supporting Figure

S3). We find an out-of-plane mode at a calculated frequency of 784 cm$^{-1}$, that in a strict backscattering geometry is not visible (see supporting Figure S4). Non-normal incidence from an objective with large NA or surface morphology may however relax this condition and consequently lead to the observation of additional modes.

**Conclusions**

In conclusion, we reported on the development of a Raman-optimized substrate and on an optimized mapping approach that allowed us to acquire detailed Raman spectroscopic information of GNRs. The RO-substrates were integrated into a device-type sample geometry which allow us to monitor the GNRs' quality during device fabrication with high signal-to-noise ratio and minimal damage to the GNRs. Finally, we investigated the GNRs' low-frequency modes, the SLM, RBLM and its overtones and matched their frequency and mode profiles to first-principles calculations. Overall, both the RO-substrate and the optimized measurement approach allowed unprecedented insight into the low-frequency modes of GNRs and demonstrated their usefulness in monitoring GNR quality upon device fabrication.

**Experimental Methods**

*On-surface synthesis of AGNRs.* 9-AGNRs were synthesized from 3',6'-diiodo-1,1':2',1''- terphenyl (DITP)[35], 7-AGNRs from 10,10'-dibromo-9,9'-bianthryl (DBBA)[1] and the 5-AGNRs were synthesized from an isomeric mixture of 3,9-diiodoperylene and 3,10-diiodoperylene (DIP) (details on monomer synthesis will be published elsewhere) as the precursor monomers. The GNRs were grown on a vicinal single crystal Au(788). First, Au(788) was cleaned in ultra-high vacuum by two sputtering/annealing cycles: 1 kV Ar+ for 10 minutes followed by annealing at 420 °C for 10 minutes. In a next step, the precursor monomer was sublimed onto the Au surface from a quartz crucible heated to 70 °C (DITP) or 200°C (DBBA and DIP), with the substrate held at room temperature. In order to activate the polymerization reaction, both 7- and 9-AGNRs were heated up to 200 °C (0.5 K/s) with a 10 minute holding time. Subsequently, samples were annealed at 400 °C (0.5 K/s with a 10 minute holding time) in order to form the GNRs via cyclodehydrogenation. For the synthesis of the 5-AGNRs a slow annealing (0.2 K/s) was carried up to 225 °C.

*Substrate transfer of AGNRs.* AGNRs were transferred from their growth substrate Au(788) to the RO-substrates by an electrochemical delamination method. First, PMMA was spin coated (2500 rpm for 90 seconds, 4 layers) on GNR/Au, to act as a support layer during the transfer, followed by a 10 minutes curing step at 80 °C. In a next step the PMMA at the edges of the Au (788) crystal was removed after UV-exposure (80 minutes) followed by 3 minutes development in water/isopropanol. By removing the PMMA from the crystal's edges, the delamination time was reduced to 45-60 seconds. The electrochemical cell was mounted using a carbon rod as anode, the PMMA/GNRs/Au as the cathode and 1M NaOH as electrolyte. By applying a voltage of 5 V (current ~0.2 A) between anode and cathode, hydrogen bubbles are formed at the interface of PMMA/GNRs and Au resulting in the delamination of the PMMA/GNR-layer. After delamination, the PMMA/GNR layer was cleaned for 5 minutes in purified water before being transferred to the target substrate. In a next step, the PMMA/GNRs/substrate stack was annealed for 10 minutes at 80 °C followed by 20 minutes at 110 °C to increase the adhesion between the target substrate and the PMMA/GNR layer. Finally, the PMMA was dissolved in acetone (15 minutes) and the final GNR/substrate rinsed with ethanol and ultrapure water.

*Raman spectroscopy.* Raman spectra were acquired using a WITec Alpha 300 R confocal Raman microscope in backscattering geometry with a 50x and (100x) objectives, NA=0.55 (0.9). For aligned GNRs the linear polarization of the exciting lasers was adjusted parallel to the GNRs. The backscattered light was detected without an analyzing polarizer and coupled to one of two spectrometers: a 300 mm lens-based spectrometer with gratings of 600 g/mm or 1800 g/mm equipped with a cooled EM-CCD for measurements with 488 nm and 532 nm excitation, and a 400 mm lens-based spectrometer with gratings of 300 g/mm or 1200 g/mm and a cooled deep-depletion CCD for 785 nm excitation.

The laser wavelength, power and integration time were optimized for each type of GNR and substrate to maximize the signal and keep the intensity loss as discussed above well below 10% for all settings. Unless stated otherwise, the samples were mounted in a home-built vacuum chamber at a pressure below $10^{-2}$ mbar, mounted on a piezo stage for scanning. A polynomial background was subtracted from the raw spectra unless otherwise stated.

*First-principle calculations.* We performed DFT-calculations and calculated the Raman intensities by using VASP for energy and force calculations[31–33] in conjunction with the Phonopy program package for the calculation of phonon modes and frequencies[34]. Projector-augmented-wave pseudopotentials, a plane-wave cutoff of 600 eV, and the Perdew–Burke–Ernzerhof exchange-correlation functional were used in VASP. Raman intensities were calculated using in-house utility codes and the finite difference method[36] where the frequency-dependent dielectric matrix was also calculated via DFT[37].

**Author Contributions**

**Acknowledgments**

This work was supported by the Swiss National Science Foundation under Grant No 20PC21_155644, the European Union's Horizon 2020 research and innovation program under grant agreement number 785219 (Graphene Flagship Core 2), the Office of Naval Research (N00014-18-1-2708) and the NCCR MARVEL funded by the Swiss National Science Foundation (51NF40-182892). Raman scattering modeling used resources at the Center for Nanophase Materials Sciences, which is a DOE Office of Science User Facility. Part of the computations was performed using resources of the Center for Computational Innovation at Rensselaer Polytechnic Institute. X. Y. W., T. D., A. N., and K. M. acknowledge the support by the Max Planck Society. This work was partially funded by the FET open project QuIET (no. 767187). M.P. acknowledges funding by the EMPAPOSTDOCS-II program, which is financed by the European Union's Horizon 2020 research and innovation program under the Marie Sklodowska-Curie grant agreement number 754364. J.O. and O.B. acknowledge technical support from the Binning and Rohrer Nanotechnology Center (BRNC), J.O. and M.C. thank Erwin Hack for ellipsometry measurements, and Sascha Martin and Heinz Breitenstein for technical support.

**References**

1. Cai, J., Ruffieux, P., Jaafar, R., Bieri, M., Braun, T., Blankenburg, S., Muoth, M., Seitsonen, A.P., Saleh, M., Feng, X., Müllen, K., and Fasel, R. (2010) Atomically precise bottom-up fabrication of graphene nanoribbons. *Nature*, **466** (7305), 470–473.
2. Cai, J., Pignedoli, C.A., Talirz, L., Ruffieux, P., Söde, H., Liang, L., Meunier, V., Berger, R., Li, R., Feng, X., Müllen, K., and Fasel, R. (2014) Graphene nanoribbon heterojunctions. *Nat Nano*, **9** (11), 896–900.
3. Gröning, O., Wang, S., Yao, X., Pignedoli, C.A., Borin Barin, G., Daniels, C., Cupo, A., Meunier, V., Feng, X., Narita, A., Müllen, K., Ruffieux, P., and Fasel, R. (2018) Engineering of robust topological quantum phases in graphene nanoribbons. *Nature*, **560** (7717), 209–213.
4. Ruffieux, P., Cai, J., Plumb, N.C., Patthey, L., Prezzi, D., Ferretti, A., Molinari, E., Feng, X., Müllen, K., Pignedoli, C.A., and Fasel, R. (2012) Electronic Structure of Atomically Precise Graphene Nanoribbons. *ACS Nano*, **6** (8), 6930–6935.

5. Martini, L., Chen, Z., Mishra, N., Barin, G.B., Fantuzzi, P., Ruffieux, P., Fasel, R., Feng, X., Narita, A., Coletti, C., Müllen, K., and Candini, A. (2019) Structure-dependent electrical properties of graphene nanoribbon devices with graphene electrodes. *Carbon*, **146**, 36–43.

6. Son, Y.-W., Cohen, M.L., and Louie, S.G. (2006) Energy Gaps in Graphene Nanoribbons. *Phys. Rev. Lett.*, **97** (21), 216803.

7. Talirz, L., Ruffieux, P., and Fasel, R. (2016) On-Surface Synthesis of Atomically Precise Graphene Nanoribbons. *Adv. Mater.*, **28** (29), 6222–6231.

8. Corso, M., Carbonell-Sanromà, E., and Oteyza, D.G. de (2018) Bottom-Up Fabrication of Atomically Precise Graphene Nanoribbons, in *On-Surface Synthesis II*, Springer, Cham, pp. 113–152.

9. Talirz, L., Söde, H., Dumslaff, T., Wang, S., Sanchez-Valencia, J.R., Liu, J., Shinde, P., Pignedoli, C.A., Liang, L., Meunier, V., Plumb, N.C., Shi, M., Feng, X., Narita, A., Müllen, K., Fasel, R., and Ruffieux, P. (2017) On-Surface Synthesis and Characterization of 9-Atom Wide Armchair Graphene Nanoribbons. *ACS Nano*, **11** (2), 1380–1388.

10. Kimouche, A., Ervasti, M.M., Drost, R., Halonen, S., Harju, A., Joensuu, P.M., Sainio, J., and Liljeroth, P. (2015) Ultra-narrow metallic armchair graphene nanoribbons. *Nature Communications*, **6**, 10177.

11. Fairbrother, A., Sanchez-Valencia, J.-R., Lauber, B., Shorubalko, I., Ruffieux, P., Hintermann, T., and Fasel, R. (2017) High vacuum synthesis and ambient stability of bottom-up graphene nanoribbons. *Nanoscale*, **9** (8), 2785–2792.

12. Barin, G.B., Fairbrother, A., Rotach, L., Bayle, M., Paillet, M., Liang, L., Meunier, V., Hauert, R., Dumslaff, T., Narita, A., Müllen, K., Sahabudeen, H., Berger, R., Feng, X., Fasel, R., and Ruffieux, P. (2019) Surface-synthesized graphene nanoribbons for room-temperature switching devices: substrate transfer and ex-situ characterization. *ACS Appl. Nano Mater.*

13. Llinas, J.P., Fairbrother, A., Borin Barin, G., Shi, W., Lee, K., Wu, S., Choi, B.Y., Braganza, R., Lear, J., Kau, N., Choi, W., Chen, C., Pedramrazi, Z., Dumslaff, T., Narita, A., Feng, X., Müllen, K., Fischer, F., Zettl, A., Ruffieux, P., Yablonovitch, E., Crommie, M., Fasel, R., and Bokor, J. (2017) Short-channel field-effect transistors with 9-atom and 13-atom wide graphene nanoribbons. *Nature Communications*, **8** (1), 633.

14. Bennett, P.B., Pedramrazi, Z., Madani, A., Chen, Y.-C., Oteyza, D.G. de, Chen, C., Fischer, F.R., Crommie, M.F., and Bokor, J. (2013) Bottom-up graphene nanoribbon field-effect transistors. *Applied Physics Letters*, **103** (25), 253114.

15. Tuinstra, F., and Koenig, J.L. (1970) Raman Spectrum of Graphite. *The Journal of Chemical Physics*, **53** (3), 1126–1130.

16. Martins Ferreira, E.H. (2010) Evolution of the Raman spectra from single-, few-, and many-layer graphene with increasing disorder. *Phys. Rev. B*, **82** (12).

17. Dresselhaus, M.S., Jorio, A., Hofmann, M., Dresselhaus, G., and Saito, R. (2010) Perspectives on Carbon Nanotubes and Graphene Raman Spectroscopy. *Nano Lett.*, **10** (3), 751–758.

18. Verzhbitskiy, I.A., Corato, M.D., Ruini, A., Molinari, E., Narita, A., Hu, Y., Schwab, M.G., Bruna, M., Yoon, D., Milana, S., Feng, X., Müllen, K., Ferrari, A.C., Casiraghi, C., and Prezzi, D. (2016) Raman Fingerprints of Atomically Precise Graphene Nanoribbons. *Nano Lett.*, **16** (6), 3442–3447.

19. Vandescuren, M., Hermet, P., Meunier, V., Henrard, L., and Lambin, Ph. (2008) Theoretical study of the vibrational edge modes in graphene nanoribbons. *Phys. Rev. B*, **78** (19), 195401.

20. Gillen, R., Mohr, M., and Maultzsch, J. (2010) Raman-active modes in graphene nanoribbons. *phys. stat. sol. (b)*, **247** (11–12), 2941–2944.

21. Ma, C., Liang, L., Xiao, Z., Puretzky, A.A., Hong, K., Lu, W., Meunier, V., Bernholc, J., and Li, A.-P. (2017) Seamless Staircase Electrical Contact to Semiconducting Graphene Nanoribbons. *Nano Lett.*, **17** (10), 6241–6247.

22. Passi, V., Gahoi, A., Senkovskiy, B.V., Haberer, D., Fischer, F.R., Grüneis, A., and Lemme, M.C. (2018) Field-Effect Transistors Based on Networks of Highly Aligned, Chemically Synthesized $N = 7$ Armchair Graphene Nanoribbons. *ACS Applied Materials & Interfaces*, **10** (12), 9900–9903.

23. Senkovskiy, B.V., Pfeiffer, M., Alavi, S.K., Bliesener, A., Zhu, J., Michel, S., Fedorov, A.V., German, R., Hertel, D., Haberer, D., Petaccia, L., Fischer, F.R., Meerholz, K., van Loosdrecht, P.H.M., Lindfors, K., and Grüneis, A. (2017) Making Graphene Nanoribbons Photoluminescent. *Nano Lett.*, **17** (7), 4029–4037.

24. Spizzirri, P.G., Fang, J.-H., Rubanov, S., Gauja, E., and Prawer, S. (2008) Nano-Raman spectroscopy of silicon surfaces. *Materials Forum*, **34**, 161–166.

25. Connell, G. a. N., Nemanich, R.J., and Tsai, C.C. (1980) Interference enhanced Raman scattering from very thin absorbing films. *Appl. Phys. Lett.*, **36** (1), 31–33.

26. Bacsa, W.S., Pavlenko, E., and Tishkova, V. (2013) Optical Interference Substrates for Nanoparticles and Two-Dimensional Materials. *Nanomaterials and Nanotechnology*, **3**, 22.

27. Blake, P., Hill, E.W., Neto, A.H.C., Novoselov, K.S., Jiang, D., Yang, R., Booth, T.J., and Geim, A.K. (2007) Making graphene visible. *Applied Physics Letters*, **91** (6), 063124.

28. Abidi, I.H., Cagang, A.A., Tyagi, A., Riaz, M.A., Wu, R., Sun, Q., and Luo, Z. (2016) Oxidized nitinol substrate for interference enhanced Raman scattering of monolayer graphene. *RSC Adv.*, **6** (9), 7093–7100.

29. Liu, C., Ma, Y., Li, W., and Dai, L. (2013) The evolution of Raman spectrum of graphene with the thickness of SiO2 capping layer on Si substrate. *Appl. Phys. Lett.*, **103** (21), 213103.
30. Solonenko, D., Gordan, O.D., Milekhin, A., Panholzer, M., Hingerl, K., and Zahn, D.R.T. (2016) Interference-enhanced Raman scattering of F16CuPc thin films. *Journal of Physics D: Applied Physics*, **49** (11), 115502.
31. Kresse, G., and Hafner, J. (1993) Ab initio molecular dynamics for liquid metals. *Phys. Rev. B*, **47** (1), 558–561.
32. Kresse, G., and Furthmüller, J. (1996) Efficiency of ab-initio total energy calculations for metals and semiconductors using a plane-wave basis set. *Computational Materials Science*, **6** (1), 15–50.
33. Kresse, G., and Furthmüller, J. (1996) Efficient iterative schemes for ab initio total-energy calculations using a plane-wave basis set. *Phys. Rev. B*, **54** (16), 11169.
34. Togo, A., and Tanaka, I. (2015) First principles phonon calculations in materials science. *Scripta Materialia*, **108**, 1–5.
35. Di Giovannantonio, M., Deniz, O., Urgel, J.I., Widmer, R., Dienel, T., Stolz, S., Sánchez-Sánchez, C., Muntwiler, M., Dumslaff, T., Berger, R., Narita, A., Feng, X., Müllen, K., Ruffieux, P., and Fasel, R. (2018) On-Surface Growth Dynamics of Graphene Nanoribbons: The Role of Halogen Functionalization. *ACS Nano*, **12** (1), 74–81.
36. Liang, L., and Meunier, V. (2014) First-principles Raman spectra of $MoS_2$, $WS_2$ and their heterostructures. *Nanoscale*, **6** (10), 5394–5401.
37. Gajdoš, M., Hummer, K., Kresse, G., Furthmüller, J., and Bechstedt, F. (2006) Linear optical properties in the projector-augmented wave methodology. *Phys. Rev. B*, **73** (4), 045112.
38. Wiecha, P.R. (2018) pyGDM—A python toolkit for full-field electro-dynamical simulations and evolutionary optimization of nanostructures. *Computer Physics Communications*, **233**, 167–192.
39. Olmon, R.L., Slovick, B., Johnson, T.W., Shelton, D., Oh, S.-H., Boreman, G.D., and Raschke, M.B. (2012) Optical dielectric function of gold. *Phys. Rev. B*, **86** (23), 235147.
40. Kumar, P., Wiedmann, M.K., Winter, C.H., and Avrutsky, I. (2009) Optical properties of $Al_2O_3$ thin films grown by atomic layer deposition. *Appl. Opt., AO*, **48** (28), 5407–5412.

# Supporting Information

## Optimized substrates and measurement approaches for Raman spectroscopy of graphene nanoribbons

*Jan Overbeck*[1,2,3,#], *Gabriela Borin Barin*[1,#], *Colin Daniels*[4], *Mickael Perrin*[1], *Liangbo Liang*[5], *Oliver Braun*[1,2], *Rimah Darawish*[1,6], *Bryanna Burkhardt*[4], *Tim Dumslaff*[7], *Xiao-Ye Wang*[7,†], *Akimitsu Narita*[7], *Klaus Müllen*[7,8], *Vincent Meunier*[4], *Roman Fasel*[1,6], *Michel Calame*[1,2,3] and *Pascal Ruffieux*[1]*

[1]Empa, Swiss Federal Laboratories for Materials Science and Technology, 8600 Dübendorf, Switzerland
[2]University of Basel, Department of Physics, 4056 Basel, Switzerland
[3]University of Basel, Swiss Nanoscience Institute, 4056 Basel, Switzerland
[4]Rensselaer Polytechnic Institute, Department of Physics, Applied Physics, and Astronomy, Troy, New York 12180, United States
[5]Oak Ridge National Laboratory, Center for Nanophase Materials Sciences, Oak Ridge, Tennessee 37831, United States
[6]University of Bern, Department of Chemistry and Biochemistry, 3012 Bern, Switzerland
[7]Max Planck Institute for Polymer Research, Ackermannweg 10, 55128 Mainz, Germany
[8]Johannes Gutenberg-Universität Mainz, Institute of Physical Chemistry, 55128 Mainz, Germany

[†]current address: State Key Laboratory of Elemento-Organic Chemistry, College of Chemistry, Nankai University, Tianjin 300071, China

[#] These authors contributed equally

*corresponding authors: gabriela.borin-barin@empa.ch
jan.overbeck@empa.ch
pascal.ruffieux@empa.ch

## Supporting Note 1 – Additional calibrated spectra on Raman optimized (RO)-substrates

To assess the magnitude of the interference enhancement (IE) effect with respect to standard silicon substrates, hybrid devices with adjacent areas of both layer structures were fabricated (see inset of Figure S1a). This allows the direct comparison of Raman intensities obtained from a single scan. Figure S1a shows the enhancement obtained from a device optimized for Raman spectroscopy at an excitation wavelength $\lambda_{ex}$=785 nm (thickness of 114 nm as determined by ellipsometry). Figure S1b shows a device with a significantly thinner oxide, which is beneficial for device applications but still shows significant IE. Note, that the $SiO_2$/Si substrate itself exhibits an interference effect which results in enhancement values that differ from what is expected to an interference-free substrate.

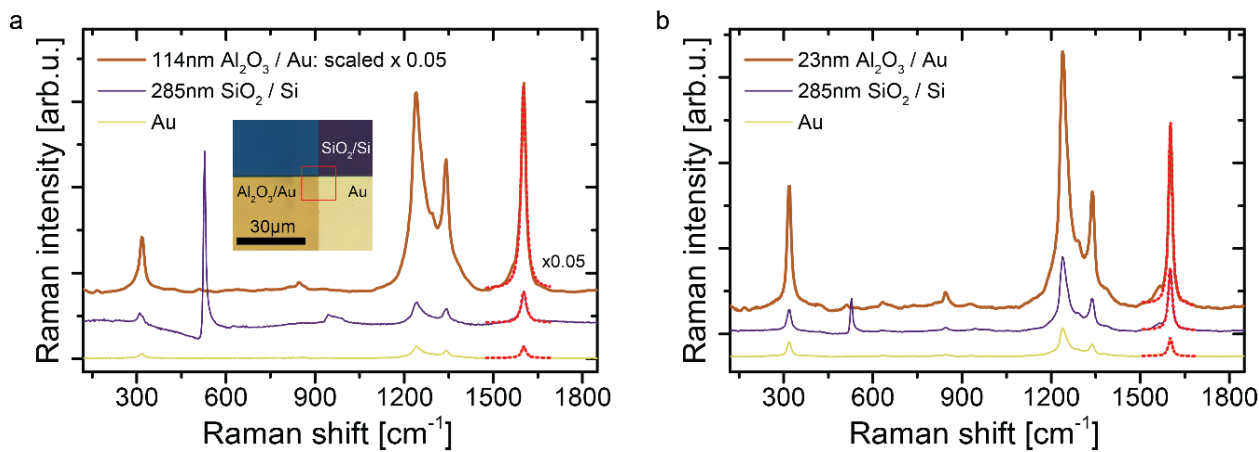

**Figure S1. Calibrated enhancement factors for selected oxide thicknesses. a** Raman enhancement of the spectrum of 9-AGNR at an oxide thickness of 114 nm optimized for 785 nm excitation. An enhancement of 120 (373) on the RO-substrate is extracted from the G-peak intensity with respect to the signal on $SiO_2$/Si (Au). Note that the enhanced spectrum had to be scaled by 1/20 to make the peaks apparent for the reference spectra. The inset show how the data are obtained from a single Raman map covering the three substrates. **b** Spectra obtained on thinner (23 nm) oxide as it may be found for atomic-layer-deposited gate oxides in devices, still showing a significant enhancement of 2.3 (9.8). Note the change in relative intensities of the silicon and GNR-signals between a and b, that reflect the variability of GNR growth and transfer. Spectra obtained in air, 100x (NA=0.9) objective.

## Supporting Note 2 – Fabrication and modelling of interference enhanced substrates

To experimentally probe the thickness-dependence of the interference effect on RO-substrates, we fabricated a sample with wedged oxide structure, following Solonenko et al. [30]. For this, we used a thick (300 nm) atomic-layer-deposited aluminum oxide layer, which was subsequently etched away while pulling the substrate out of the etchant solution (TMAH-based developer, MF-321). Figure S2a shows an optical micrograph and sketch of the resulting wedge-shaped oxide structure. Raman spectra were acquired via a map-scan for each excitation wavelength (colored outline in a). The G-mode intensity was extracted from a 10 μm averaging window along the scan direction as shown in Figure 3c of the main manuscript. The thickness was calibrated via profilometer measurements.

To model the interference enhancement we calculated the intensity of the excitation laser resulting from self-interference using the pyGDM package [38]. We employ a layer structure consisting of a $SiO_2$ substrate and a 90 nm thick layer of gold covered by aluminum oxide (Figure S2b). We use a frequency-dependent refractive index for Au [39] and a refractive index for Al$_2$O$_3$ of n$_{ox}$=1.67 [40]. This accounts for the frequency-dependent skin-depth which shifts the interference maxima to lower values compared to the simple interference model for thin layers, which predicts the first interference maximum at an oxide thickness of d$_{ox}$= λ/(4*n$_{ox}$) ≈ 118nm (80/73 nm) for 785nm (532/488 nm) excitation. Further deviations are attributed to the non-normal incidence for an objective NA = 0.55 and the effect of self-interference of Raman shifted scattered light[30].

For RO-device substrates, an oxide thickness of around 40 nm is used and the oxide layer is etched away except for dedicated areas on the source/drain contact pads (see Figure 3b of the main manuscript).

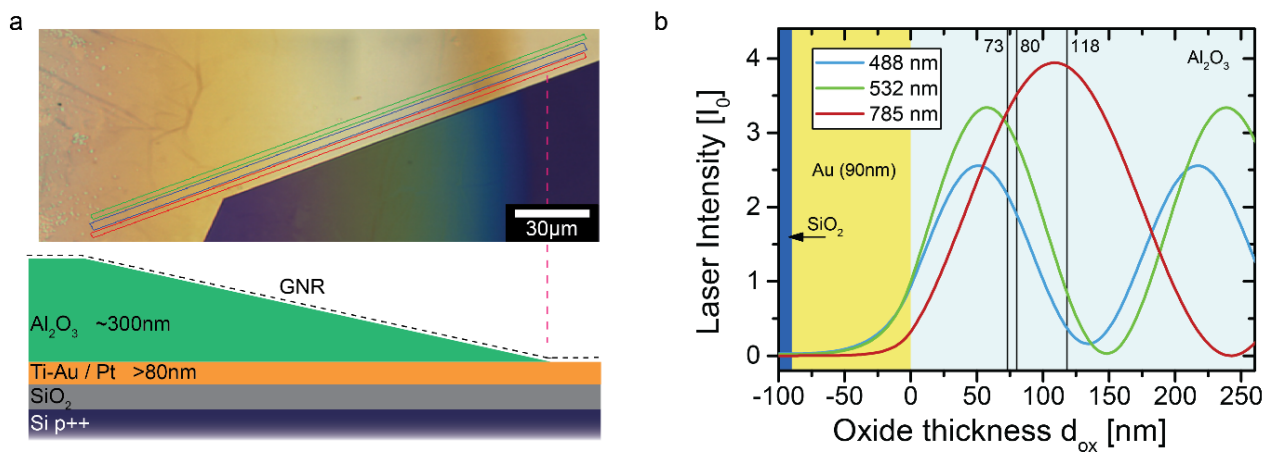

**Figure S2. Wedged oxide structure for investigation and modelling of interference enhancement. a** Substrate structure used for acquiring oxide thickness dependent spectra. **b** Normalized E$^2$ as a function of oxide thickness at the oxide surface (position of the GNR-layer). The measured Raman intensity is shown for comparison.

**Supporting Note 3 – Raman data on additional samples and normal mode displacements**

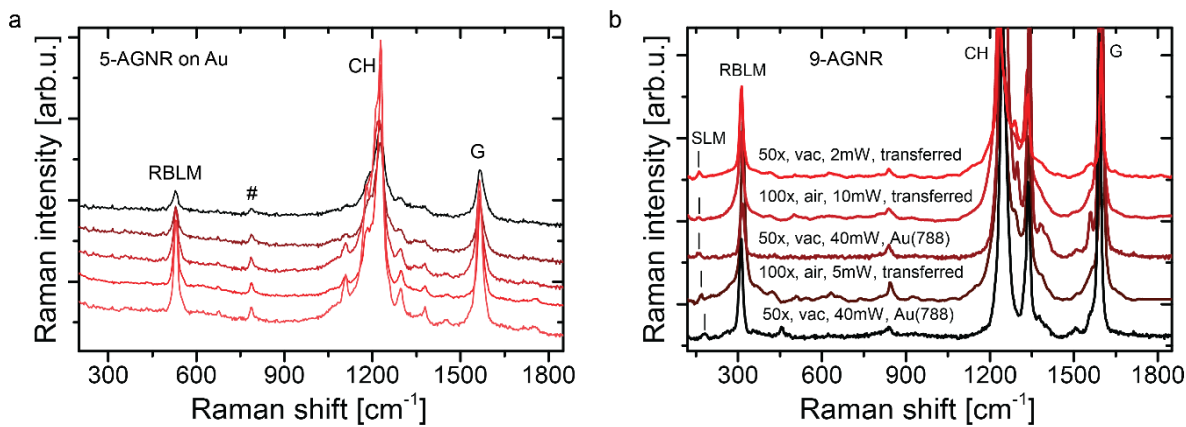

**Figure S3. Additional Spectra. a** Spectra from 5-Samples of 5-AGNR on Au growth-substrates. All spectra each constructed via averaging a large area map, exhibit mode '#'. Excitation wavelength = 785 nm, 40 mW, in vacuum. No background subtraction. **b** Raman spectra of 9-AGNR sample before and after transfer as indicated. On each sample, the SLM at 160-180 cm$^{-1}$ is clearly visible. Excitation wavelength 785 nm.

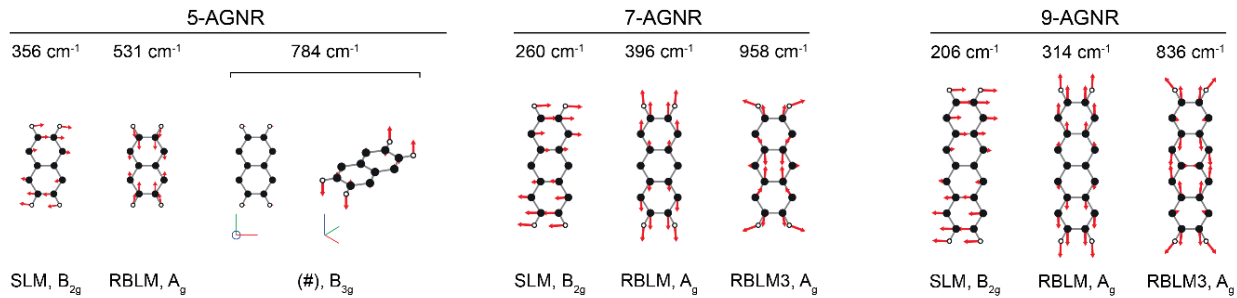

**Figure S4. Normal mode displacements for low-energy modes.** Mode frequencies and symmetries are indicated. For the 5-AGNR, the mode "#" which is experimentally observed at 789 cm$^{-1}$ is tentatively attributed to a mode with $B_{3g}$ symmetry at a Raman shift of 784 cm$^{-1}$. The out-of-plane vibration can be seen in the side-projections. It becomes Raman allowed for non-normal incidence, which can be due to the GNR-on-substrate morphology and the deviation from normal excitation and detection due to the objective NA=0.55. Normal modes for the low energy 7- and 9-AGNR are shown for comparison. For further calculations we refer to our previous work[12] .